%% file: main.tex
\documentclass{article}

\usepackage{microtype}
\usepackage{graphicx}
\usepackage{subcaption}
\usepackage{booktabs} 
\usepackage{color}

\usepackage{hyperref}

\usepackage{graphicx} 
\usepackage{amsmath}
\usepackage{amsfonts}
\usepackage{booktabs}
\usepackage{todonotes}
\usepackage{makecell}
\usepackage{multirow}
\usepackage{amssymb}
\usepackage{breakurl}

\usepackage{pifont}
\newcommand{\cmark}{\ding{51}}%
\newcommand{\xmark}{\ding{55}}%

\usepackage[font=footnotesize]{caption}

\usepackage{adjustbox}

\newcommand{\sys}{Seesaw}

\newcommand{\bhbm}{B_\text{HBM}}

\newcommand{\flops}{\text{FLOPS}}

\newcommand{\TP}{\text{TP}}
\newcommand{\PP}{\text{PP}}
\newcommand{\DP}{\text{DP}}
\newcommand{\attn}{\text{attn}}
\newcommand{\linear}{\text{linear}}

\usepackage{tikz}

\usepackage{bm}
\usepackage[accepted]{mlsys2024}

\usepackage{enumitem}
\setlist[itemize]{leftmargin=*}

\usepackage[para]{footmisc}

\begin{document}

\twocolumn[
\mlsystitle{\sys{}: High-throughput LLM Inference via Model Re-sharding}



\mlsyssetsymbol{equal}{*}

\begin{mlsysauthorlist}
\mlsysauthor{Qidong Su\footnotemark[1]\footnotemark[2]\footnotemark[3]}{}
\mlsysauthor{Wei Zhao\footnotemark[3]\footnotemark[4]}{}
\mlsysauthor{Xin Li\footnotemark[3]}{}
\mlsysauthor{Muralidhar Andoorveedu\footnotemark[3]}{}
\mlsysauthor{Chenhao Jiang\footnotemark[1]\footnotemark[2]}{}
\mlsysauthor{Zhanda Zhu\footnotemark[1]\footnotemark[2]\footnotemark[3]}{}
\mlsysauthor{Kevin Song\footnotemark[1]\footnotemark[2]}{}
\mlsysauthor{Christina Giannoula\footnotemark[1]\footnotemark[2]\footnotemark[3]}{}
\mlsysauthor{Gennady Pekhimenko\footnotemark[1]\footnotemark[2]\footnotemark[3]}{}
\end{mlsysauthorlist}



\mlsyskeywords{Large Language Model, Distributed Inference}

\vskip 0.3in

\input{content/abstract}

]

\footnotetext[1]{University of Toronto}
\footnotetext[2]{Vector Institute}
\footnotetext[3]{CentML}
\footnotetext[4]{Stanford University}





\input{content/1-intro}
\input{content/2-background}
\input{content/3-motivation}
\input{content/4-keyidea}
\input{content/5-impl}
\input{content/6-eval}

\input{content/7-related}
\input{content/8-conclusion}

\nocite{langley00}

\bibliography{ref.bib}
\bibliographystyle{mlsys2024}


\appendix
\input{content/a1-model}


\end{document}

%% file: content/abstract.tex
\begin{abstract}
To improve the efficiency of distributed large language model (LLM) inference, various parallelization strategies, such as tensor and pipeline parallelism, have been proposed. However, the distinct computational characteristics inherent in the two stages of LLM inference—prefilling and decoding—render a single static parallelization strategy insufficient for the effective optimization of both stages.
In this work, we present \emph{\sys{}}, an LLM inference engine optimized for throughput-oriented tasks. The key idea behind \sys{} is \emph{dynamic model re-sharding}, a technique that facilitates the dynamic reconfiguration of parallelization strategies across stages, thereby maximizing throughput at both phases.
To mitigate re-sharding overhead and optimize computational efficiency, we employ \emph{tiered KV cache buffering} and \emph{transition-minimizing scheduling}. These approaches work synergistically to reduce the overhead caused by frequent stage transitions while ensuring maximum batching efficiency.
Our evaluation demonstrates that \sys{} achieves a throughput increase of up to 1.78$\times$ (1.36$\times$ on average) compared to vLLM, the most widely used state-of-the-art LLM inference engine.
\end{abstract}

%% file: content/1-intro.tex
\section{Introduction}\label{sec:intro}

Large language models (LLMs), such as the LLaMA~\cite{llama} and GPT~\cite{gpt4} families, have demonstrated exceptional performance across a wide range of tasks. Beyond their prevalent use in interactive applications like chatbots~\cite{chatgpt}, LLMs are also gaining high interest in throughput-oriented offline inference workloads such as information extraction~\cite{narayan2022can}, database querying~\cite{llm-sql}, and knowledge graph processing~\cite{graphrag}. 
Unlike interactive applications where low latency is crucial, these offline inference tasks \emph{prioritize high throughput over response time}.
These offline inference workloads are widely adopted in industry~\cite{offline-bytedance, offline-aws, offline-dell, offline-snowflake}, leading MLPerf to develop benchmarks specifically for them~\cite{mlcommons_datacenter_benchmark}.
In this work, we focus on improving inference efficiency for offline, throughput-oriented LLM inference workloads.

As LLMs often exceed the memory capacity of individual GPUs, parallelization is essential for their deployment ~\cite{demystifying-dist, megatron}. Several parallelization strategies, including \emph{tensor} parallelism~\cite{megatron} and \emph{pipeline} parallelism~\cite{pipedream, gpipe}, have been proposed, each presenting distinct trade-offs in memory efficiency, inter-device communication, and computational efficiency.
Tensor parallelism distributes model weights across devices but suffers from high communication costs due to frequent all-reduce operations at each layer~\cite{scale-transformer, flux}.
The communication cost becomes particularly severe in systems connected via PCIe~\cite{dell_gpu_matrix} or with partial high-speed connections~\cite{a100-pcie}. In contrast, pipeline parallelism partitions the model into sequential stages, reducing inter-device communication by passing only activations between them. However, to enable pipelining, each data batch needs to be divided into micro-batches, leading to extra execution overheads, since every micro-batch repeatedly loads weights into the compute units (see Section~\ref{sec:parallel} for details).

\input{content/figs/motivation}

While numerous studies have proposed methods to optimize parallelization strategies for LLMs~\cite{miao2023towards, vllm, alpaserve, scale-transformer}, prior works typically rely on a single, static configuration throughout the entire generation process.
However, our findings indicate that this one-size-fits-all approach is often inefficient for \emph{throughput-oriented} LLM inference because it fails to leverage the distinct patterns between the two stages in LLM generation: the \emph{prefill} stage, where the input sequence is processed at once to produce the initial token, and the \emph{decode} stage, where subsequent tokens are generated sequentially based on prior tokens.
These two stages exhibit fundamentally different computational characteristics~\cite{llmviewer}.
During the prefill stage, multiple tokens from the input prompt are processed simultaneously, making computation and communication the dominant contributors to runtime. In contrast, the decode stage processes one token at a time for each sequence, increasing the relative time spent on weight transfer.
This difference indicates that the optimal parallelization strategy for each stage may also vary.

To illustrate the performance limitations of applying a uniform parallelization strategy for both prefill and decode, we measure the execution time of each stage under various combinations of tensor and pipeline parallelism, as shown in Figure~\ref{fig:motivation-example}.
In the \underline{prefill} stage, as the degree of tensor parallelism increases, the communication overhead increases significantly due to additional GPUs participating in all-reduce operations. As a result, tensor parallelism performs significantly worse than pipeline parallelism.
In contrast, during the \underline{decode} stage, pipeline parallelism is slower than tensor parallelism, largely due to increased weight transferring overhead caused by micro-batching required for pipelining (see Section~\ref{sec:parallel} for more details).
Therefore, we need stage-specific parallelization strategies to provide better LLM inference throughput.

An existing approach is disaggregated prefill-decode~\cite{distserve,mooncake}, which assigns prefill and decode computation to different GPU instances. The prefill instances and decode instances form a two-stage pipeline to serve inference requests.
Therefore, the overall throughput of disaggregated prefill-decode is constrained by the slower of the two stages, and balancing throughput between these two stages is essential.
The key drawback of disaggregated prefill-decode is that it can cause large amounts of pipeline bubbles under resource-constrained environments.
For example, when deploying a 70B model on 8$\times$40GB GPUs, even the most balanced configuration results in a 6$\times$ difference in throughput between the prefill and decode stages. In this setup, the decode stage operates at one-sixth the throughput of the prefill stage, resulting in a significant bottleneck at the prefill stage that slows down the entire system (see Section~\ref{sec:why-not-distserve} for details).

To address these challenges, we present \emph{\sys{}}, a high-throughput LLM inference engine that dynamically reconfigures parallelization strategies between the prefill and decode stages. The key idea behind \sys{} is \textbf{model re-sharding}, a novel technique that dynamically re-partitions model weights and KV cache \footnote{The tensors cached for each sequence’s decoding steps.} between prefill and decode stages. 
By tailoring parallelization strategies to the distinct computational demands of each stage, \sys{} reduces communication overhead during the prefill stage, while enhancing memory efficiency in the decode stage, resulting in a substantial increase in overall throughput.

\input{content/figs/schedule}
However, the overhead associated with model re-sharding can be high due to frequent transitions between prefill and decode. To maximize throughput, existing systems typically adopt prefill-prioritized scheduling~\cite{orca,vllm}, which interleaves prefill and decode stages across batches to achieve continuous batching. Yet, as illustrated in Figure~\ref{fig:schedule}(a), integrating this approach with model re-sharding can result in significant overhead due to frequent transitions between prefill and decode. On the other hand, decode-prioritized scheduling~\cite{faster-transformer} completes all decode steps for a batch before proceeding to the next, resulting in lower re-sharding overhead. However, as depicted in Figure~\ref{fig:schedule}(b), this method suffers from low resource utilization due to smaller batch sizes. 

To overcome this constraint and achieve both minimal re-sharding overhead and large batch size, we propose two synergetic techniques: \textbf{tiered KV cache buffering} and \textbf{transition-minimizing scheduling}. 
Tiered KV cache buffering leverages CPU memory as auxiliary storage for the KV cache, enabling \sys{} to store the KV cache for a large number of prefill requests.
Transition-minimizing scheduling reduces re-sharding overhead by minimizing the number of transitions to the decode stage. \sys{} transitions from prefill to decode only after the CPU KV cache is full. During decoding, the large number of KV cache in the CPU buffer enables Seesaw to perform decode with large batch sizes, and thus enabling high throughput.
As depicted in Figure~\ref{fig:schedule}(c), this approach maintains the maximal batch size during the decode stage, while significantly reducing the frequency of stage transitions, thereby minimizing re-sharding overhead. 
Additionally, to mitigate the overhead of KV cache transfers between CPU and GPU, \sys{} employs asynchronous pipelining to overlap data transfers with computation.

In summary, we make the following contributions.
\begin{itemize}
\setlength\itemsep{-0.1em}
\item We identify and quantitatively analyze the different preferences for parallelisms in the prefill and decode stages of \emph{throughput-oriented} LLM inference tasks. Our analysis comprehensively accounts for data movement, computation, and communication costs.
\item We propose \emph{dynamic model re-sharding}, a novel technique that dynamically reconfigures the parallelization strategies for prefill and decode stages. We address the challenge of transition overhead in model re-sharding with continuous batching by introducing \emph{tiered KV cache buffering} and \emph{transition-minimizing scheduling}. Based on these techniques, we implement \sys{}, a high-throughput offline inference system that optimizes parallelization strategies for each LLM inference stage.
\item We conduct a comprehensive evaluation of \sys{} across a variety of workloads and hardware configurations. Our results show \sys{} achieves an average speedup of 1.36$\times$ and a throughput improvement of up to 1.78$\times$ compared to the state-of-the-art LLM inference engines.
\end{itemize}

%% file: content/figs/motivation.tex
\begin{figure}[t]
\centering
\subfloat[Prefill]{
    \centering
    \includegraphics[width=0.46\linewidth]{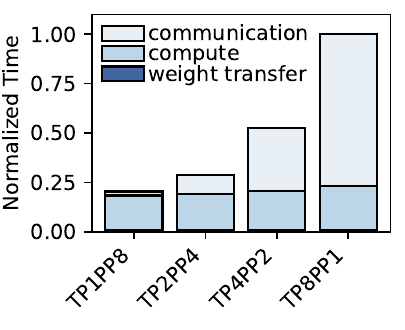}
    \label{fig:motivation-prefill}
}\hfill
\subfloat[Decode]{
    \centering
    \includegraphics[width=0.46\linewidth]{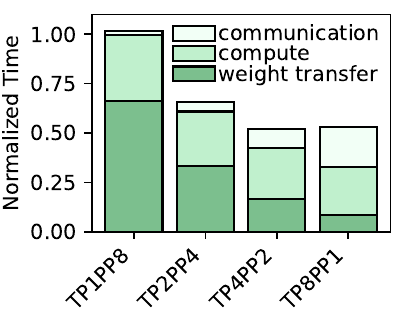}
    \label{fig:motivation-decode}
}
\caption{Breakdown of execution time for the prefill and decode stages for LLaMA2-13B inference on 8 L4 GPUs (The global batch size is 16. Pipeline parallelism further divides the data into micro-batches of size $16/\PP$ to  fully utilize pipelining).}
\label{fig:motivation-example}
\end{figure}

%% file: content/figs/schedule.tex
\begin{figure}
    \centering
    \includegraphics[width=0.95\linewidth, trim={0.2cm 0.2cm 0.2cm 0.2cm}]{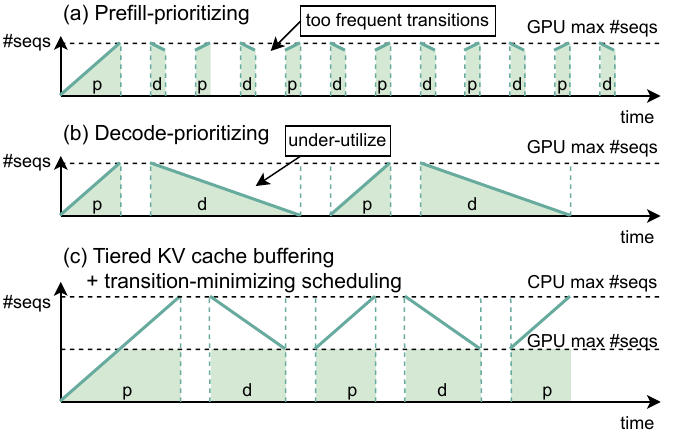}
    \caption{Different scheduling policies considering transition overhead. Decoding throughput is positively correlated with the number of sequences in GPU memory (the maximal batch size), which is highlighted as light green area.}
    \label{fig:schedule}
\end{figure}

%% file: content/2-background.tex
\section{Background}

\begin{figure*}[t]
    \centering
    \includegraphics[width=\linewidth]{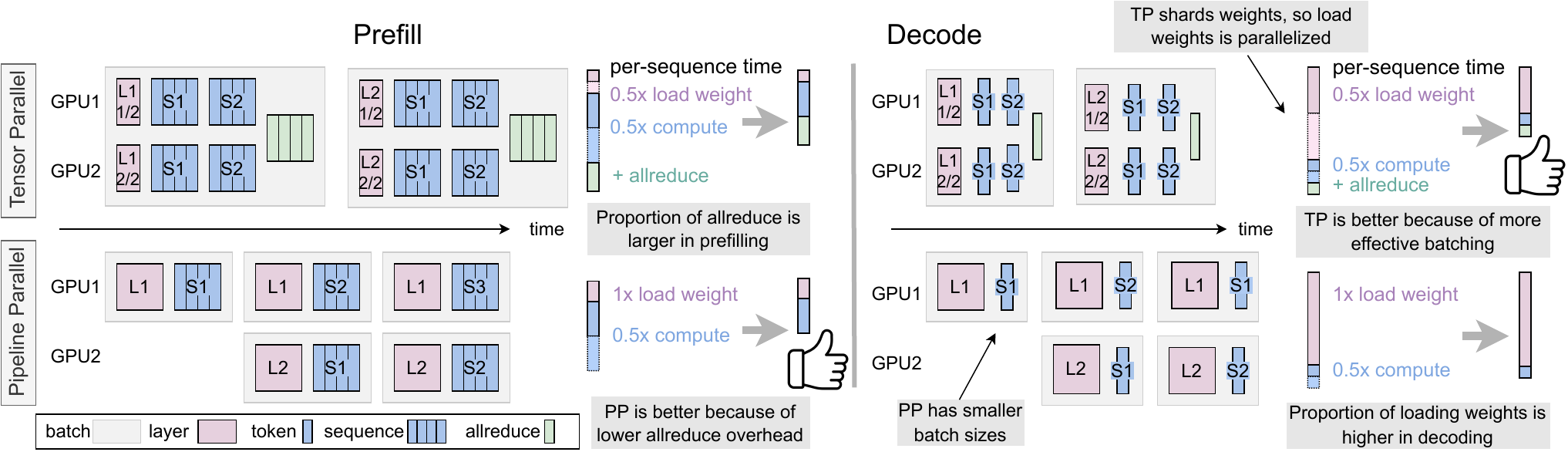}
    \caption{Different effects of tensor and pipeline parallelisms on prefilling and decoding. Tensor parallelism incurs all-reduce overhead, which has a higher percentage in prefilling, therefore pipeline parallelism is better for prefilling. Conversely, pipeline parallelism splits batches into smaller micro-batches, which leads to more forward passes and repetitive loading weights, which is insufficient in decoding.}
    \label{fig:motivation}
\end{figure*}

\subsection{LLM Inference}

\paragraph{Transformer Architecture.} 
Modern large language models are based on the transformer architecture~\cite{attention}, which typically consists of multiple identical decoder layers~\cite{chatgpt}.
Each layer includes several linear layers and an attention layer.
The weights of the linear layers account for the majority of the model’s parameters.

\paragraph{Auto-regressive Generation.}
LLM inference follows an auto-regressive paradigm~\cite{auto-regression}, which takes an input prompt and generates a sequence of output tokens. 
This process is divided into two stages: prefilling, which processes the input tokens, and decoding, which generates a token per step. 
These stages exhibit distinct computational properties~\cite{distserve, llmviewer}.
Prefilling processes the prompt that are typically hundreds to thousands of tokens long. 
The computation and communication costs, both of which scale with the number of tokens, dominate the runtime during this stage.
Since the cost of loading weights is amortized over a larger set of tokens, the overall performance is primarily bound by compute and/or communication.
In contrast, Decoding processes only the newly generated tokens in each auto-regressive step and has comparatively smaller computation in each step.
Therefore the cost for loading the weight data from off-chip memory to computation units has a relatively higher percentage. 
In each generation step, the intermediate tensors $K$ and $V$ in each attention operator can be cached for reuse in the future generation, which is called Key-value cache (\emph{KV cache})~\cite{scale-transformer}. While being able to accelerate computation, it occupies a substantial amount of GPU memory, which is proportional to the total number of tokens.

\subsection{LLM Inference Optimization}\label{sec:llm-opt}

\paragraph{Parallelism.}
As the size of LLMs grows, the memory capacity on a single GPU becomes insufficient.
Consequently, various techniques are developed to partition models onto multiple GPUs~\cite{alpa}.
These parallelization strategies can be classified as (1) inter-operator, which places different operators or layers across multiple GPUs, overlapping them with pipelining (known as \emph{Pipeline parallelism}, PP)~\cite{gpipe, pipedream, alpaserve}, and (2) intra-operator, which partitions different dimensions of tensors involved in computation, including data parallelism~\cite{preble}, tensor parallelism~\cite{megatron}, etc. 
\emph{Data parallelism} duplicates models on different devices and dispatches requests among them. 
\emph{Tensor parallelism} shards model weights and each device performs a portion of the computation, then aggregates these partial results to produce the final output.

\paragraph{Batching.}
Batching more tokens in a single forward pass increases inference efficiency by, for example, amortizing the time required to load model weights~\cite{flexgen,turbotransformers}. However, its effectiveness differs between the prefilling and decoding stages~\cite{llmviewer,fastdecode,sarathi}. In decoding, where weight-loading overhead occupies a larger portion of the runtime, batching significantly boosts throughput by effectively amortizing this overhead.
Conversely, in the prefilling stage, batching has a less pronounced impact since the token count in input prompts is generally sufficient to keep the process compute-bound. Overall, larger batch sizes yield higher throughput, though the maximum batch size is limited by available GPU memory, as it requires additional space for activations and the KV cache.

\paragraph{Continuous Batching and Scheduling.}
Continuous batching is an essential optimization for throughput-oriented LLM inference~\cite{orca,vllm}. 
By batching multiple sequences at the token level, it allows the system to onboard new sequences and clear the KV cache of completed sequences at any generation step. 
This approach enables prefill-prioritizing scheduling, which removes sequences as they finish, frees up their KV cache, and eagerly schedules the prefilling of new sequences whenever GPU memory becomes available. 
This strategy maximizes the number of concurrent sequences being processed, resulting in higher throughput.
Another alternative is to use \emph{decode-prioritizing} scheduling, which minimizes the frequency of transitions.
Instead of scheduling to prefilling eagerly, this approach waits until all sequences in a batch have finished decoding before initiating the next round of prefilling.
However, this scheduling policy results in suboptimal decoding throughput~\cite{sarathi-serve}.

%% file: content/3-motivation.tex
\section{Motivation and Analysis}

In this section, we provide an in-depth analysis of two key observations we identify from Figure~\ref{fig:motivation-example} in Section~\ref{sec:intro}: (1) Tensor parallelism often exhibits significantly worse performance than pipeline parallelism during the prefill stage due to its substantial communication overhead; (2) Pipeline parallelism tends to fall short in the decode stage owing to the considerable weight loading overhead it incurs. We then argue that a dynamic parallelization strategy is essential to attain optimal performance across both stages. 

Given the importance of batching in throughput-oriented tasks, it can be useful to consider how different parallelization strategies impact the \emph{maximum batch size}, rather than assuming batch size as a tunable parameter, as is often done in online-serving contexts such as DistServe~\cite{distserve} and Sarathi-serve~\cite{sarathi-serve}.

\subsection{Parallelism Analysis}\label{sec:parallel}

\paragraph{Observation 1: Tensor parallelism incurs substantial communication overhead during the prefill stage.}
In Tensor parallelism, each device performs a part of computation and aggregate the partial result. The activations at each layer are synchronized across all GPUs using all-reduce operations. The overhead associated with this operation can be quantified as:
$$
\frac{\#\text{tokens} \times \text{activation size} }{\text{all-reduce bandwidth}},
$$
where \textit{all-reduce bandwidth} refers to the rate of data transfer during all-reduce operations, calculated as the size of the tensor being all-reduced divided by the all-reduce runtime.

As the degree of tensor parallelism increases, the proportion of execution time of all-reduce operations grows substantially. This growth is attributed to two main factors. First, while model weights are partitioned, activations in tensor parallelism remain fully replicated across GPUs, leading to a constant activation size regardless of the degree of tensor parallelism.
Second, all-reduce bandwidth decreases as the number of GPUs grows, due to more complex communication schemes.
Therefore, increasing the degree of tensor parallelism not only fails to reduce the traffic of all-reduce operations but further limits the communication bandwidth, resulting in escalated communication overhead.
This issue is particularly pronounced in the prefill stage, where a large number of tokens are processed simultaneously, making communication overhead the primary bottleneck. Thus, tensor parallelism tends to perform worse than pipeline parallelism due to its large communication overhead.

\paragraph{Observation 2: Pipeline parallelism suffers from significant weight transferring overhead in the decode stage.}
Pipeline parallelism distributes model layers sequentially across devices, with each device responsible for processing a set of consecutive layers before passing the output to the next device.
Due to the auto-regressive nature of LLM inference, a sequence cannot enter the pipeline until its preceding token is generated. As a result, at any given time step, a sequence can appear in only one stage of the pipeline, making the batches processed by each device \emph{mutually exclusive}. However, the total number of sequences that the pipeline can handle at a time, referred to as the \emph{global batch size}, is constrained by the size of KV cache. Given the mutual exclusion of batches at each device, pipeline parallelism can process only approximately $1/\PP$ of the global batch per forward pass. We denote this reduced batch size in pipeline parallelism as the \emph{micro-batch size}.

Dividing batches into micro-batches increases the number of LLM forward passes required to process the same amount of requests. Specifically, a pipeline parallelism degree of $\PP$ necessitates $\PP$ times more forward passes for a given global batch. This repeated execution degrades inference performance, as model weight matrices must be loaded from global memory repeatedly. This inefficiency is especially significant in the decode stage, where weight-loading overhead accounts for a substantial portion of total execution time. As a result, pipeline parallelism generally underperforms relative to tensor parallelism in the decode stage due to the amplified weight loading overhead.

\paragraph{Discussion on Data Parallelism.}
Unlike tensor and pipeline parallelism, which distribute the model across devices, data parallelism distributes the data while duplicating the model.
While data parallelism has minimal communication overhead, it has two key disadvantages: (1) the volume of weight transferring is higher by the number of duplicates compared to tensor parallelism; and (2) it occupies more GPU memory, reducing the available space for the KV cache and thus limiting the maximum batch size resulting in lower throughput. 
%
Data parallelism can be applied orthogonally alongside both tensor and pipeline parallelism. We do not dynamically adjust data parallelism, which will be explained in Section~\ref{sec:no-dp}.

\paragraph{Conclusion: No one-size-fits-all}
When comparing these three parallelism strategies for high-throughput LLM inference, a key observation is that prefilling and decoding stages benefit from different parallelism approaches. This difference arises from the distinct characteristics of each stage, as illustrated in Figure~\ref{fig:motivation}.
Tensor parallelism is preferred for decoding due to its ability to efficiently accelerate weight matrix loading. However, it incurs significant communication overhead, as it requires all-reduce operations at each layer. In contrast, pipeline and data parallelism have much lower communication overhead, making them preferable for prefilling. However, their decoding throughput is limited by inefficient batching and additional weight-loading overhead.

To quantitatively analyze the trade-offs across different parallelisms, we model the average runtime per sequence (the inverse of throughput) as follows. Derivations and further details are provided in the Appendix~\ref{app:model}.
\begin{equation}\label{eqn:tput}
\small
T \propto \frac{T_{dm}^\linear}{\TP} + \frac{T_{dm}^\attn + T_{comp}}{\DP\cdot\TP\cdot\PP} + \frac{T_{comm}(\TP)}{\PP\cdot\DP}\notag
\end{equation}
Here $T_{dm}^\linear$ represents data movement for linear layers (primarily model weights), $T_{dm}^\attn$ represents data movement for attention layers (primarily KV cache)
, $T_{comp}$ represents computation time, $T_{comm}$ represents communication time. Note that $T_{comm}$ is a monotonically increasing function with respect to $\TP$,  as all-reduce operations require more time as $\TP$ increases. 

Tensor parallelism can effectively accelerate loading model weights, which is $T_{dm}^\linear$, while pipeline and data parallelism cannot. On the other hand, pipeline and data parallelism effectively reduce the overhead of communication, while tensor parallelism contrarily increases the communication overhead. In prefilling, $T_{dm}^\linear$ is negligible, and $T_{comm}$ becomes larger, so pipeline and data parallelisms are more preferred, while in decoding, $T_{dm}^\linear$ occupies a larger proportion so tensor parallelism is more advantageous.

\subsection{Why not Disaggregate Prefilling and Decoding?}\label{sec:why-not-distserve}
Spatially disaggregating prefilling and decoding with separate hardware resources, as done in online serving systems such as DistServe~\cite{distserve} and MoonCake~\cite{mooncake}, is one approach to separately select parallelization strategies for prefilling and decoding. Sequences are first processed by the devices dedicated for prefilling before being transferred to decoding devices.

\input{content/figs/distserve}

However, there are two obstacles when applying prefill-decode disaggregation to purely throughput-oriented scenarios. 
First, since the overall throughput is bound by the slower stage, the throughput of prefilling and decoding needs to be matched by adjusting the devices allocated for each stage. However, it can be impractical in resource-constrained scenarios. As shown in Figure~\ref{fig:distserve-mismatch}, to deploy a 70B model (which takes 140GiB memory for model weights) on eight 40GiB GPUs, there is only one disaggregation strategy, that is four GPUs for prefilling and four for decoding\footnote{At least four GPUs (160 GiB memory) are needed to fit the model weights.}. However, it causes severe throughput mismatch where prefilling has more than 6$\times$ higher throughput than decoding.
Second, disaggregation duplicates the model weights similarly to data parallelism, bringing similar drawbacks, such as limited KV cache space and increased weight transfer. As a result, decoding throughput with four GPUs is only 15\% of that with eight GPUs.

In conclusion, although disaggregation allows for selecting different parallelization strategies for each stage, the throughput mismatch between stages and limited resources allocated to each can lead to suboptimal performance. This calls for a method that offers flexibility in parallelization while maximizing hardware resource utilization.

%% file: content/figs/distserve.tex
\begin{figure}[t]
    \centering
    \captionsetup[subfigure]{labelformat=empty, skip=0cm}
    \subfloat[]{
        \includegraphics[width=0.9\linewidth]{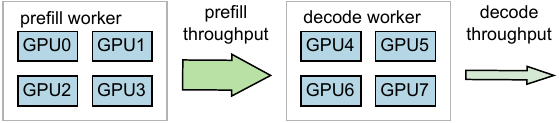}
    }\\
    \subfloat[]{
        \includegraphics[width=0.9\linewidth]{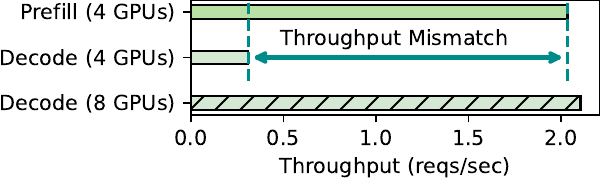}
    }
    \caption{An example of spatially disaggregating prefilling and decoding has a restricted search space. Deploying a 70B model on eight 40GiB GPUs allows only one disaggregation strategy: four GPUs for prefilling and four for decoding. However, this causes severe throughput mismatch between the two stages.
    }
    \label{fig:distserve-mismatch}
\end{figure}

%% file: content/4-keyidea.tex
\section{\sys{}: Key Ideas}


\subsection{Dynamic Model Re-sharding}\label{sec:why-reshard}
Observing that prefilling and decoding have distinct preferences for parallelism, we propose a technique called \emph{dynamic model re-sharding}. This technique enables the selection of different parallelism strategies for each stage and automatically transitions between them. This approach expands the configuration space, allowing for separate optimization of the two stages, potentially improving overall throughput compared to using a single configuration.  In the following paragraphs, we denote the parallelization strategy used in prefilling as $c_p$ and that in decoding as $c_d$.

\begin{figure}[t]
    \centering
    \includegraphics[width=\linewidth]{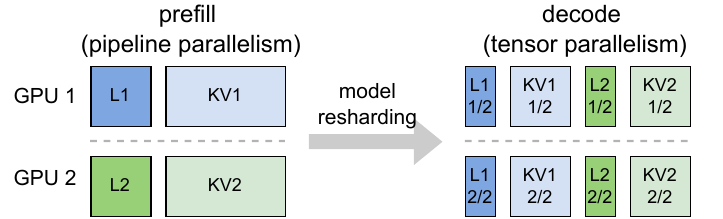}
    \caption{Model weights and KV cache need to be re-sharded when switching between different parallelism.}
    \label{fig:reshard-overview}
\end{figure}

To support transitions between different parallelization configurations, the cluster must rearrange the data stored on each device to align with the new parallelism which involves both model weights and KV cache, as illustrated in Figure~\ref{fig:reshard-overview}.  
In \sys{}, model weights are re-sharded by reloading the required shards from CPU memory, and KV cache re-sharding is performed through CPU shared memory.

The inter-device movement of tensors incurs overhead. To mitigate this re-sharding cost, we design an asynchronous pipeline to overlap data transfer with computation, as detailed in Section~\ref{sec:async-pipeline}.

\paragraph{Discussion on data parallelism.}\label{sec:no-dp}
Unlike switching between tensor and pipeline parallelism, adjusting the degree of data parallelism alters the proportion of GPU memory allocated to model weights versus KV cache. This adjustment increases system complexity or necessitates additional data movement between the CPU and GPU. Therefore, we only dynamically adjust tensor and pipeline parallelism.

\subsection{Tiered KV Cache Buffering and Transition-minimizing Scheduling}\label{sec:swap}
\begin{figure}[t]
\centering
\includegraphics[width=0.95\linewidth]{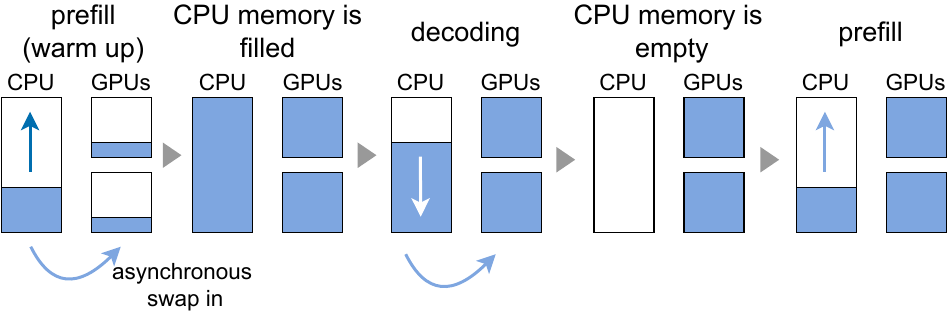}
\caption{Tiered KV cache buffering and transition-minimizing scheduling, and the change of KV cache occupancy.}
\label{fig:reshard-loop}
\end{figure}

\begin{figure}[t]
\centering
\includegraphics[width=\linewidth]{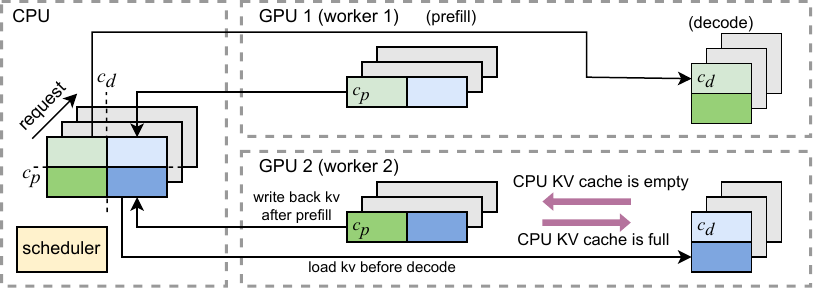}
\caption{KV cache re-sharding is completed during swapping, leveraging CPU shared memory.}
\label{fig:repartition}
\end{figure}

\paragraph{Challenge: Transition Overhead.}
In practice, dynamic model resharding encounters an obstacle of transition overhead, which is amplified by the widely-used continuous batching and prefill-prioritizing scheduling.
Prefill-prioritizing scheduling eagerly schedules new prefilling tasks, causing frequent transitions between the two stages.
As a result, directly applying model re-sharding with this interleaved prefill-decode scheduling policy would introduce significant re-sharding overhead.
On the other hand, decode-prioritizing scheduling minimizes the frequency of transitions but results in suboptimal decoding throughput.
Other compromise solutions involve setting a threshold-based approach for managing the prefill-decode transition~\cite{cheng2024slice}.
However, they still involve a trade-off between reducing transition overhead and maximizing decoding throughput.

To address this problem, we propose 1) \emph{tiered KV cache buffering}, which leverages CPU memory offloading 2) \emph{transition-minimizing} scheduling policy. These two synergistic techniques prevent frequent stage transitions and maintain a high decoding throughput.

Tiered KV cache buffering uses CPU memory as auxiliary storage for the KV cache, enabling the pre-computation of a large batch of prefilling consecutively. During the prefill stage, the generated KV cache is offloaded to CPU KV cache storage, freeing it from the limitations of GPU memory space. During decoding, continuous batching runs as normal, except that new sequences are on-boarded by swapping in its KV cache from the CPU memory.

Transition-minimizing scheduling controls the transition to only happen when the CPU KV storage is either full or empty.
During prefill, once the CPU KV cache storage is fully utilized, re-sharding is triggered, and the cluster transitions to decoding.
During decoding, GPUs continue processing requests and loading KV cache from CPU memory, keeping GPU KV cache fully utilized for high decoding throughput.
When the entire CPU KV cache has been transferred to GPU memory, the cluster switches back to prefilling.
The whole process is illustrated in Figure~\ref{fig:reshard-loop}.

KV cache re-sharding occurs throughout this process.
As illustrated in Figure~\ref{fig:repartition}, in a multi-GPU setup, the CPU KV cache storage is shared among all GPUs.
During swap-out, each GPU pushes its shard (based on $c_p$) of the generated KV cache to the shared CPU storage, where these shards collectively form the complete KV cache. During swap-in, each GPU retrieves its required KV shard (based on $c_d$) from the shared storage. We implement the shared KV cache using shared memory of the operating system.

%% file: content/5-impl.tex
\section{System Design and Implementation}


\subsection{Scheduler-worker Architecture}
In order to support dynamically switching parallelization configurations for prefilling and decoding, we build \sys{}, a new LLM inference engine designed for high-throughput LLM inference. 
The overall architecture of \sys{} follows a single-scheduler, multi-worker design.
The scheduler manages all generation requests, organizes them into batches, and sends instructions to the workers. To fully utilize pipelining, each decoding step processes $1/\PP$ of the sequences in GPU KV storage. Once a batch is formed, it is sent to workers through shared queues.
Each worker is responsible for controlling a single GPU and maintains a task queue to receive and execute instructions sequentially.
This architecture facilitates the implementation of asynchronous features, such as pipeline parallelism and the asynchronous pipeline for tiered KV cache buffering.

\subsection{Asynchronous Pipeline}\label{sec:async-pipeline}

\begin{figure}[t]
    \centering
    \includegraphics[width=\linewidth]{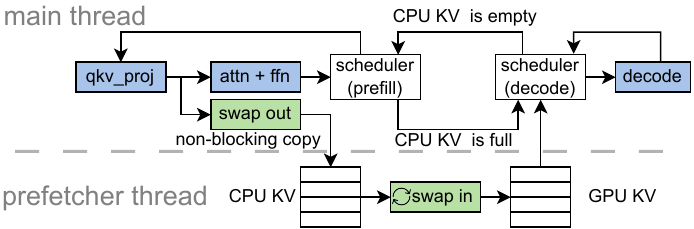}
    \caption{Async pipeline of \sys{}: Swap-in overlaps with prefill computation, while swap-out occurs in a separate asynchronous prefetcher thread.}
    \label{fig:async}
\end{figure}

While re-sharding and tiered KV cache buffering offer substantial benefits, they also introduce new overhead related to moving model weights and KV cache. 
The overhead of reloading model weights remains constant relative to batch size, allowing it to be amortized with larger batches. 
In contrast, swapping the KV cache incurs overhead proportional to batch size, making it harder to amortize. Fortunately, these overheads can be mitigated through computation-communication overlap. We implement an asynchronous pipeline to overlap KV cache transfer with ongoing computation, as illustrated in Figure~\ref{fig:async}.

\paragraph{Overlap swap-out with computation.}
The KV cache generated during the prefilling stage is not used until decoding begins, allowing the KV cache swap-out to overlap with other computations during prefilling.
Although CPU-GPU data transfer is relatively slow due to PCIe bandwidth limitations, it can still be overlapped with computation, given the high FLOPS involved in prefilling.

In practice, CPU-GPU data transfer can only overlap with computation when using pinned memory, but shared memory cannot be pinned~\cite{pytorch_discussion_pin_memory}. To address this, we split the transfer into two stages: GPU to pinned memory (overlapped with computation) and then pinned to shared memory, which is a host-side operation that also runs concurrently with GPU kernels.

\paragraph{Asynchronous swap-in.}
We implement swap-in using a background thread called the \emph{prefetcher} on each worker, operating in a fully asynchronous paradigm.
The prefetcher is controlled directly by the scheduler and runs independently of the main thread, whether the main thread is handling prefilling or decoding. 
In each iteration, the scheduler creates new prefetching tasks when there are free slots in the GPU KV store. 
Once the prefetcher completes moving the KV cache for certain sequences, it notifies the scheduler via a shared queue, allowing those sequences to be scheduled for decoding tasks later.
As long as the output length is not too short, the swap-in can also be well overlapped.

\paragraph{Bandwidth-aware KV cache layout.}
The data layout of the KV cache significantly impacts the bandwidth efficiency of data movement.
There are two common layouts for storing KV cache: \textit{(seq\_len, num\_heads, head\_dim)} (NHD) and \textit{(num\_heads, seq\_len, head\_dim)} (HND).
NHD is less optimal for memory access because tensor parallelism shards the KV cache along the $H$ dimension (number of heads), which is the second-to-last dimension, leading to more noncontiguous memory access. Therefore, we use the HND layout for storing the KV cache in CPU memory. 


%% file: content/6-eval.tex
\section{Evaluation}

In this section, we evaluate the performance of \sys{} under a variety of hardware configurations and workloads.

\subsection{Experiment Settings}

\paragraph{Hardware.}
We use three types of GPUs: NVIDIA A10, L4, and A100. The A10 and L4 are deployed on AWS EC2 instances (\texttt{g5.48xlarge} and \texttt{g6.48xlarge}~\cite{aws-ec2}), and the A100 is used on GCP~\cite{gcp-a100}. GPU specifications are listed in Table~\ref{tab:hardware}. The PCIe connection for each GPU is PCIe 4.0 8$\times$, providing 16 GiB/s bandwidth~\cite{pcie4}, while NVLink~\cite{nvlink} offers a bandwidth of 600 GiB/s. Additionally, we allocate 80 GiB of CPU memory per GPU.

\paragraph{Model.}

We use three different LLMs with different sizes: (1) a 15B variety of LLaMA3~\cite{llama3-15b}; (2) CodeLLaMA-34B~\cite{codellama}; (3) LLaMA2-70B~\cite{llama2}.
They all use Grouped Query Attention (GQA)~\cite{gqa}.
For brevity, we refer to them as 15B, 34B, and 70B, respectively, in the following sections.
We use float16 as the data type. 

\paragraph{Workload.}

We use two different datasets in our evaluation, namely \texttt{sharegpt}~\cite{sharegpt} and \texttt{arxiv-summarization}~\cite{arxiv-dataset}. They correspond to two different distributions of workload. \texttt{sharegpt} is a dataset of chatting history, so its input and output have comparable lengths, while \texttt{arxiv-summarization} dataset is a summarization dataset where inputs are much longer than outputs. The characteristics of these two datasets are shown in Figure~\ref{fig:dataset}. We sample 2000 requests from the \texttt{sharegpt} dataset and 500 requests from \texttt{arxiv-summarization} and also use constant-length workloads in Section~\ref{sec:sensitivity}.
Since \sys{} is purely throughput-oriented, we measure the end-to-end throughput as the metrics.

\paragraph{Baselines.} We use vLLM 0.5.4~\cite{vllm} as the baseline. It is the most widely used open-source LLM serving engine with wide support for different parallelisms.
We also directly use the vLLM's model implementation for a straightforward comparison. 
SGLang~\cite{sglang} and DeepSpeed-FastGen~\cite{deepspeed-fastgen} do not support pipeline parallelism. 
TensorRT-LLM~\cite{trt-llm} is not included in the comparison because it uses a similar scheduling policy as vLLM, and vLLM demonstrates comparable performance~\cite{vllm2024} in throughput-oriented tasks. The techniques proposed in \sys{} can also be applied to modifying TensorRT-LLM.

We enable chunked prefill and tune the chunk size for vLLM to get the optimal throughput, following the practice of Sarathi-serve~\cite{sarathi-serve}. Otherwise, suboptimal chunk sizes would cause severe throughput degradation.

\input{content/tables/hardware}
\input{content/figs/datasets}

\subsection{End-to-end Throughput on PCIe Systems}\label{sec:eval-pcie}

\input{content/figs/e2e-l4-a10}

First, we measure the end-to-end throughput of \sys{}. We sweep over all available single parallelism configurations for vLLM and show the result of the best configuration. We use four GPUs for the 15B model, and eight GPUs for the 34B and 70B models. The result is shown in Figure~\ref{fig:e2e-tput}, with the used parallelism labeled above each bar.

On A10, compared with the highest single parallelism baseline, \sys{} achieves a geometrically average speedup of 1.45$\times$, with up to 1.78$\times$ speedup. On L4, \sys{} achieves a geometrically average speedup of 1.29$\times$, with up to 1.52$\times$ speedup. The overall average speedup is 1.36$\times$.
The speedup is more significant on A10, because A10 has better single GPU performance than L4, while they have similar PCIe inter-connection bandwidth, causing a higher percentage of communication overhead.

\subsection{Speedup Breakdown: An Example}
\begin{figure}[t]
    \centering
    \includegraphics[width=\linewidth]{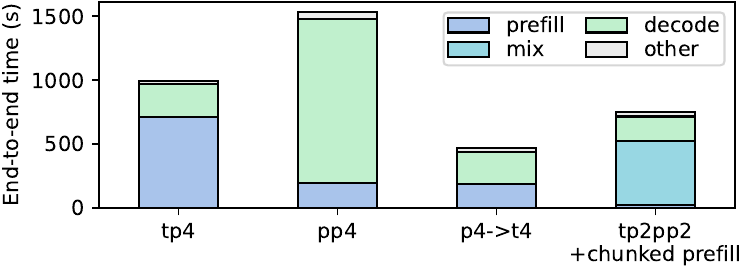}
    \caption{Speedup breakdown. ``mix'' represents batches containing both prefilling and decoding when chunked prefill is enabled. We disable chunked prefill for TP4 and PP4 in order to show the reference prefilling and decoding time. TP2PP2 with chunked prefill is the optimal parallelism for vLLM.}
    \label{fig:breakdown}
\end{figure}

Figure~\ref{fig:breakdown} illustrates how \sys{} merges the advantages of different parallelisms.
Using CodeLLaMA34B on the \texttt{arxiv-summarization} dataset with four A10 GPUs as an example, we measured the runtime of each stage. TP4 is optimal for decoding but significantly slower for prefilling, while PP4 excels at prefilling but is slower during decoding. \sys{} uses a mixed parallelism strategy, applying PP4 for prefilling and TP4 for decoding, achieving performance comparable to the best configuration for each stage.

Compared to the optimal single parallelism configuration (TP2PP2) with chunked prefill, \sys{} is  still faster because (1) chunked prefill does not piggy-back all decoding steps, leaving some purely decoding steps, and (2) chunked prefill with TP2PP2 is slower than prefilling with PP4.

\subsection{End-to-end Throughput on A100}\label{sec:eval-a100}

\input{content/figs/e2e-a100}

\paragraph{Speedup on A100 + NVLink} The NVLink interconnection across A100 GPUs significantly reduces the all-reduce overhead and further scales tensor parallelism. Usually, tensor parallelism alone is enough to achieve optimal performance when there are no more than four GPUs.
Nevertheless, there is still a noticeable percentage of all-reduce overhead in prefilling when tensor parallelism scales beyond four GPUs. \sys{} can still provide speedup in this case. As shown in Figure~\ref{fig:e2e-a100}, \sys{} still achieves a 13\% throughput increase over vLLM for the \texttt{sharegpt} dataset on LLaMA3-70B on eight A100s.

\paragraph{Speedup on A100 + PCIe} Besides A100 SXM with NVLink inter-connection, there is also another version of A100 that is inter-connected with PCIe links, where \sys{} can achieve noticeable speedup. As shown in Figure~\ref{fig:e2e-a100}, \sys{} provides 46\% speedup on \texttt{arxiv-summarization} and 30\% speedup on \texttt{sharegpt}.
\sys{} brings the performance of the A100 PCIe version much closer to the performance level of the NVLink version. vLLM gets roughly 60\% throughput on A100 PCIe compared with A100 SXM, while \sys{} boosts it up to 82\% -- 89\%.

\subsection{Sensitivity Study}\label{sec:sensitivity}

\begin{figure}[t]
    \centering
    \includegraphics[width=\linewidth,]{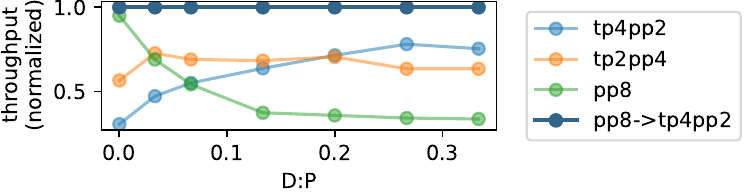}
    \caption{Throughput of various parallelization strategies with different ratios between output and input lengths ($D:P$), measured on 70B model and eight A10 GPUs. }
    \label{fig:sensitivity-pd}
\end{figure}

\paragraph{Ratio between Input and Output Length}
The speedup of \sys{} depends on the ratio between the input and output length, or $P:D$. Model re-sharding has the opportunity to provide speedup when prefilling and decoding have balanced time. To investigate to what extent model re-sharding would be effective, we measure the throughput of various parallelization strategies on synthesized datasets with uniform lengths and different $P:D$ ratios. We fix the input length as 3000 and vary the output length.

As shown in Figure~\ref{fig:sensitivity-pd}, PP8 achieves the highest throughput during prefilling, while TP4PP2 excels in decoding. When the output length equals one (prefilling only), \sys{} and PP8 show similar throughput, and TP4PP2 performs worse due to high communication overhead. As output length increases, the inefficiency of PP in decoding outweighs its advantage in prefilling, causing PP8's throughput to drop rapidly. There is a range where TP2PP4 becomes optimal before decoding dominates the runtime and TP4PP2 takes over as the fastest. Nonetheless, \sys{} achieves the highest overall throughput across all data points. In real scenarios with variable input and output lengths, \sys{} is even more advantageous due to its adaptive capabilities.

\paragraph{Inter-connection Bandwidth}
\begin{figure}[t]
    \centering
    \includegraphics[width=\linewidth]{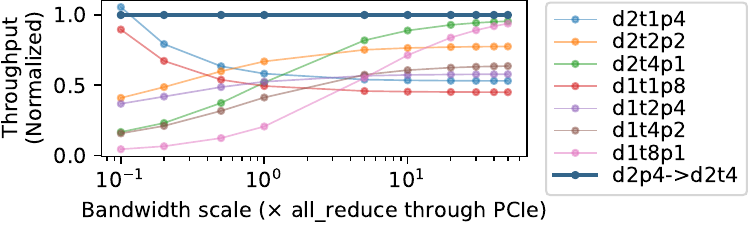}
    \caption{Projected throughput of various parallelization strategies with different inter-connection bandwidth, measured and traced on 34B model and eight A10 GPUs.}
    \label{fig:sensitivity-nw}
\end{figure}
The effectiveness of \sys{} also depends on the inter-connection bandwidth. We investigate this by measuring the runtime and tracing all-reduce operations of running \texttt{arxiv-summarization} and 34B model on eight A10s. We then mutate the all-reduce time to project the end-to-end throughput with different inter-connection bandwidths. As shown in Figure~\ref{fig:sensitivity-nw}, when the inter-connection bandwidth is slow (for example, among geographically distributed devices~\cite{petals}), pipeline parallelism is optimal; when the bandwidth is very high, tensor parallelism is optimal. The throughput of  \sys{} is superior to fixed parallelization strategies on a wide range from $0.1\times$ to $50\times$ of PCIe bandwidth.

%% file: content/tables/hardware.tex
\begin{table}[t]
\centering
\small
\caption{GPU hardware specification}
\resizebox{\columnwidth}{!}{
\begin{tabular}{c|c c c c}
\toprule
GPU Model & Memory Size & \makecell{Memory\\Bandwidth} & FLOPS & NVLink\\
\midrule
A10 & 24 GiB & 600 GiB/s & 125T & \xmark\\
L4 & 24 GiB & 300 GiB/s & 121T & \xmark\\
A100 & 40 GiB  & 1,555 GiB/s & 312T & \cmark\\
\bottomrule
\end{tabular}
}
\label{tab:hardware}
\end{table}

%% file: content/figs/datasets.tex
\begin{figure}[t]
    \centering 
    \subfloat[arxiv-summarization]{
        \includegraphics[width=0.47\linewidth]{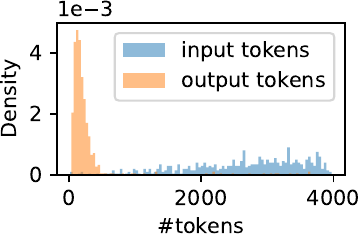}
    }
    \subfloat[ShareGPT]{
        \includegraphics[width=0.47\linewidth]{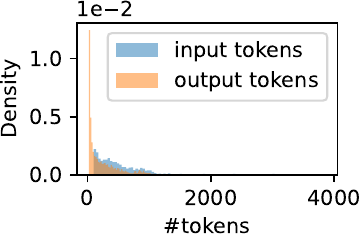}
    }
    \caption{Input and output length distributions of the  datasets}
    \label{fig:dataset}
\end{figure}

%% file: content/figs/e2e-l4-a10.tex
\begin{figure}[t]
    \centering
    \subfloat[End-to-end Throughput on A10]{
        \includegraphics[width=0.95\linewidth]{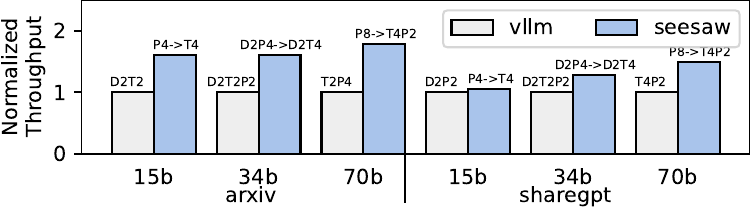}
        \label{fig:e2e-tput-a10}
    }\\
    \subfloat[End-to-end Throughput on L4]{
        \includegraphics[width=0.95\linewidth]{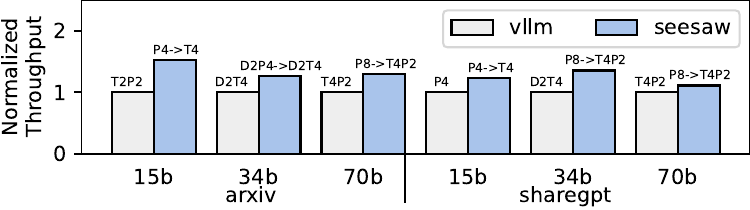}
        \label{fig:e2e-tput-l4}    
    }
    \caption{End-to-end throughput comparison on PCIe systems. The used parallelization strategies are labelled above each bar. Labels such as ``P4 $\rightarrow$ D4'' represent the parallelization strategies for prefilling and decoding respectively in \sys{}.}
    \label{fig:e2e-tput}
\end{figure}

%% file: content/figs/e2e-a100.tex
\begin{figure}[t]
    \centering
    \includegraphics[width=\linewidth]{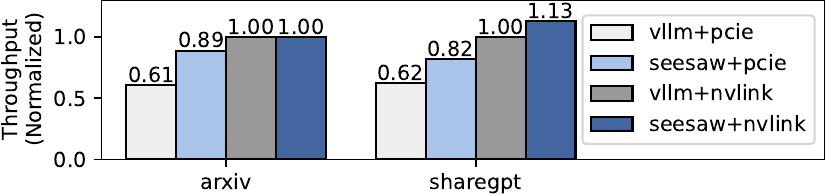}
    \caption{Throughput comparison on A100.}
    \label{fig:e2e-a100}
\end{figure}

%% file: content/7-related.tex
\section{Related Work}

\subsection{Heterogenity between Prefilling and Decoding}

Due to the different computational characteristics between prefilling and decoding leading to under-utilization of hardware resources, prior research has investigated two directions to address this problem, namely disaggregating or merging the two stages. 
Disaggregation places prefilling and decoding onto different devices to avoid their interference while merging processes prefilling and decoding in one batch.
\paragraph{Disaggregate Prefill and Decoding}
DistServe~\cite{distserve} proposed placing prefilling and decoding on different devices to prevent interference and leverage different characteristics of the two stages. 
Mooncake~\cite{mooncake} uses similar through a distributed KV cache pool.
P/D-Serve~\cite{pd-serve} uses the device-to-device network to transfer the KV cache between prefill and decode devices.
Splitwise~\cite{splitwise} proposes using different GPU models for the two stages.
TetriInfer~\cite{tetriInfer} further disaggregates different downstream tasks to avoid interference.
These works are designed for online serving while \sys{} focuses on offline inference. 
Moreover, they are usually designed for large clusters.

\paragraph{Merge Prefill and Decode}
Chunked prefill, as proposed by SplitFuse~\cite{deepspeed-fastgen}, Sarathi~\cite{sarathi}, and Sarathi-serve~\cite{sarathi-serve}, splits long prompts in the prefilling stage into smaller chunks, combining them with decoding steps to strike a balance between data movement and computation and reduce pipeline bubbles in pipeline parallelism.
However, determining the optimal chunk size is challenging. A chunk size that's too large results in excessive decode-only steps, closely resembling traditional prefill-decode scheduling. Conversely, a chunk size that’s too small reduces kernel efficiency. 

\subsection{Parallel and Distributed LLM Inference}
Aside from tensor parallelism, pipeline parallelism, and data parallelism discussed in Section~\ref{sec:llm-opt}, there are also other types of parallelisms, such as sequence parallelism (SP)~\cite{sp, ringattn, infinitellm, striped, xue2024longvila} and fully sharded data parallelism (FSDP)~\cite{pytorch-fsdp,zero}.
Sequence parallelism is especially designed for long sequence lengths, and is orthogonal with our work. FSDP requires frequently transferring weight matrices across GPUs, thus mainly used in training.

HexGen~\cite{hexgen}, LLM-PQ~\cite{llm-pq}, Helix~\cite{helix} investigate parallelisms in heterogeneous clusters.
Intra-device parallelism leverages overlapping functions using different resources within each device, including NanoFlow~\cite{nanoflow} and Liger~\cite{liger}.
Petals~\cite{petals} explores LLM inference in geographically distributed setups, employing pipeline parallelism to minimize communication costs. SpotServe~\cite{spotserve} runs LLM inference on preemptible instances.

\subsection{Offloading in LLM Inference}
Offloading is a widely used technique to run LLM applications in resource-constrained scenarios~\cite{zero-offload}.
FlexGen~\cite{flexgen} swaps tensors across GPU memory, CPU memory, and disks.
Fiddler~\cite{fiddler}, HeteGen~\cite{hetegen}, PowerInfer~\cite{powerinfer} and FastDecoder~\cite{fastdecode} perform part of computation in CPU, which require CPUs with strong compute capability or external CPU nodes connected with high-bandwidth networking.
Instinfer~\cite{instinfer} offloads computation to Computational Storage Drives.

%% file: content/8-conclusion.tex
\section{Conclusion}
This paper proposes \sys{}, a high-throughput LLM inference engine, to address the inefficiencies of fixed parallelization by selecting different parallelization strategies for the prefilling and decoding stages and switching between them using model re-sharding. It uses tiered KV cache buffering to minimize re-sharding overheads. Our experiments show that \sys{} outperforms widely-used open-source inference engines, with a throughput increase of 1.06-1.78$\times$ and an average throughput improvement of 1.36$\times$. These results highlight \sys{}'s effectiveness and adaptability.

%% file: content/a1-model.tex
\section{Performance Model}\label{app:model}

In this section, we examine the trade-offs of various parallelism strategies by developing an analytical performance model. We break down the model's inference time into multiple components and analyze the impact of each parallelism type on these components. The results reveal that the proportion of these components differs across workloads, resulting in distinct scaling behaviors for each parallelism strategy. Table~\ref{tab:notations} lists the notations used in our analysis. We assume the data type is float16.

\input{tables/notation}

\subsection{Runtime Break-Down}
The runtime of each decoding layer can be divided into three components: 1) data movement ($T_{dm}$) from GPU global memory (HBM) to compute units, which includes transferring weights ($T_{dm}^\linear$) and KV cache ($T_{dm}^\attn$), 2) computation $T_{comp}$, including $T_{comp}^\linear$ and $T_{comp}^\attn$, and 3) communication cost $T_{nw}$ ($nw$ stands for network), primarily arising from the all-reduce operation in tensor parallelism. Based on the roof-line model, the runtime of each layer can be approximated as $T_L = \max(T_{dm}^\linear, T_{comp}^\linear) + \max(T_{dm}^\attn, T_{comp}^\attn) + T_{nw}$. 

\paragraph{Data Movement.} The runtime of data movement can be approximated as transferred data volume divided by the bandwidth, which is the HBM bandwidth for GPUs. For linear layers, the transferred data is mostly weight matrices, of which the size is $2W$ bytes, which is constant. For attention layers, the transferred data is most the Q, K, and V matrices, which is $2bs(h_q+2h_{kv})d$ bytes in prefilling and $4bsh_{kv}d$ in decoding.

\paragraph{Compute.} The computation time can be approximated as the number of floating operations (FLOPs) divided by the number of floating operations per second of the hardware (FLOP/s). For linear layers, the FLOPs is proportional to the weight parameters times the number of tokens, which is $2Wbs$ in prefilling and $2Wb$ in decoding. For attention layers, most operations come from computing the attention score, which is approximated as $bh_qs^2d^2$ in prefilling and $2bh_qsd^2$ in decoding.

\paragraph{Communication.} The communication cost mostly comes from the all-reduce operation in tensor parallelism.
It can be modeled as the transferred data volume divided by the bandwidth. 
We denote it as $T_{nw}(\TP)$, and approximate it as $b \cdot A / B_{ar}(\TP)$ where $A$ is the size of the activation of one request within a batch and $B_{ar}(\TP)$ is the all-reduce bandwidth. $T_{nw}(\TP)$ is monotonically increasing with $\TP$ as additional GPUs and more replicas of activations are added to all-reduce.
We omit the peer-to-peer communication over in pipeline parallel since it is negligible compared to the all-reduce operation of tensor parallel.

\input{content/tables/model}

\subsection{Batching Analysis}
Batching is critical in decoding. It significantly affects the latency and throughput.
Batch size represents how many requests are processed in one forward pass, and larger batch sizes can amortize the cost of transferring weights, thus improving the throughput.

\paragraph{Global and micro-batch size.}
In distributed inference such as multi-GPU settings, we define the \emph{global batch size} $b$ as the number of requests being actively processed by the whole cluster. It is a tunable hyper-parameter that represents the overall workload of the system. It is bounded by the \emph{maximal batch size}, which is determined by the memory budget. 
On the other side, the \emph{micro batch size} is defined at the device level as the batch size processed during each forward pass. Tensor parallelism does not affect the batch size while DP and PP shrink the micro batch size.

\begin{figure}[t]
    \centering
    \includegraphics[width=\linewidth]{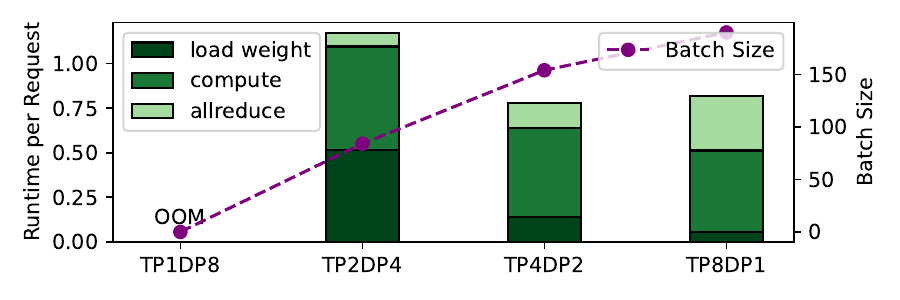}
    \caption{How data parallelism affects the decoding throughput. Data parallelism has minimal communication overhead but suffers from caused by inefficient memory access caused by duplicating model weights. Model duplicates occupy more GPU memory, leaving less space for KV cache and smaller batch sizes.
    With more data parallelism, the overhead of loading data from GPU global memory to compute units significantly increases.}
    \label{fig:motivation-dp}
\end{figure}

\subsection{Parallelism Analysis}\label{sec:parallelism}

We consider three types of parallelism: data parallelism, tensor parallelism, and pipeline parallelism, and denote their degree of parallelism as $\DP$, $\TP$, and $\PP$ respectively.

\paragraph{Tensor parallelism} can accelerate both data moving ($T_{dm}^\linear$ and $T_{dm}^\attn$ are reduced to $1/\TP$) and computation $T_{comp}$ (reduced to $T_{comp}/\TP$), at the cost of all reduce overhead $T_{nw}$.
\paragraph{Data parallelism} distributes the global batch size $b$ onto $\DP$ micro-batches processed in parallel. The model is duplicated so $T_{dm}^\linear$ remains unchanged. $T_{dw}^\attn$, $T_{comp}^\linear$, $T_{comp}^\attn$, $T_{nw}$ are reduced as the batch size is smaller.  Due to the need to duplicate model weights, the GPU memory left for the KV cache is smaller. The spare space for KV cache on each GPU is $M_{kv} = M - \frac{2LW}{TP\cdot PP}$. 
The maximal batch size is 
\begin{align*}
b_{\max} = \DP\cdot\frac{M_{kv} \cdot \TP \cdot \PP}{4 Lh_{kv} d  s} = \DP\cdot\frac{M\cdot \TP \cdot \PP - 2LW}{4Lh_{kv}ds}
\end{align*}
While TP and PP can super-linearly scale the batch size, DP can only linearly scale the batch size. The trade-off between limited batch sizes and reduced communication overhead is shown in Figure~\ref{fig:motivation-dp}.

\paragraph{Pipeline parallelism} distributes different layers to different devices, and each device will have $L/\PP$ layers. It cannot reduce single-request latency but is more suitable for throughput-oriented scenarios as it introduces less communication overhead. However, it is not the ultimate answer of high-throughput applications because of an important observation that 
\emph{pipeline parallelism harms maximal batch size.} A tricky nuance is that given a batch size $b$, pipeline parallelism can only process $b/\PP$ of them simultaneously in order to utilize and pipeline all $\PP$ GPUs, which is harmful to batching. If the workload is not uniformly distributed across GPUs, there will be bubbles, or in the worst case, some GPUs might be idle.  When the pipeline is fully and stably pipelining, each time the last pipeline stage finishes its $L/\PP$ layers of forward pass, a micro-batch of $b/\PP$ will be finished.

\paragraph{Throughput.} The micro-batch size on each GPU is $b/(\PP \cdot \DP)$. The total runtime of generating one micro batch with size $b/(\PP\cdot \DP)$ on one DP replica (or more specifically, the time of the last pipeline stage finishing a micro-batch) is  
\begin{align*}
T_\text{stage} &= \frac{L}{\PP}\cdot\left[\max(\frac{T_{dm}^\linear}{\TP}, \frac{T_{comp}^\linear}{\DP\cdot\TP\cdot\PP}) + \right.\\
&+\left.\frac{\max(T_{dm}^\attn, T_{comp}^\attn)}{\DP\cdot\TP\cdot\PP} + \frac{T_{nw}(\TP)}{\PP\cdot\DP}\right]
\end{align*}
The throughput (number of processed requests per unit time) is $b/\PP/T$. For simplicity, we calculate the inverse of it as
\begin{align}\label{eqn:tput}
\small
&\text{throughput}^{-1} = \frac{T_\text{stage}}{b/\PP} = \frac{L}{b}\cdot\left[\max(\frac{T_{dm}^\linear}{\TP}, \frac{T_{comp}^\linear}{\DP\cdot\TP\cdot\PP})\right.\notag\\
& + \left.\frac{\max(T_{dm}^\attn, T_{comp}^\attn)}{\DP\cdot\TP\cdot\PP} + \frac{T_{nw}(\TP)}{\PP\cdot\DP}\right]
\end{align}

If we approximate the roof-line model with a simplified additional model, this expression can be simplified as:
\begin{align}
\small
\text{throughput}^{-1} \propto \frac{T_{dm}^\linear}{\TP}
 + \frac{T_{comp}^\linear + T_{dm}^\attn + T_{comp}^\attn}{\DP\cdot\TP\cdot\PP} + \frac{T_{nw}(\TP)}{\PP\cdot\DP}
\end{align}




%% file: tables/notation.tex
\begin{table}[ht]
\footnotesize
\caption{Notations}
\adjustbox{max width=\linewidth}{%
\centering
    \begin{tabular}{c|c || c | c}
    \toprule
        $N$ & number of output tokens & $T_{dm}^{\linear}$ & time of moving weights \\
        $r$ & number of requests & $T_{dm}^{\attn}$ & time of moving kvcache\\
        $b$ & (global) batch size & $T_c^\linear$ & time of computation\\
        $s$ & average sequence length & $T_c^\attn$ & computation of attention \\
        $h_q$ & number of heads & $T_{nw}$ & time of communication\\
        $d$ & head dimension & $h_{kv}$ & number of KV heads\\
        $L$ & number of layers & $W$ & \#parameters of one layer \\
        $\PP$ & pipeline parallel degree & $\DP$ & data parallel degree\\
        $\TP$ & tensor parallel degree\\
    \bottomrule
    \end{tabular}
}
\label{tab:notations}
\end{table}

%% file: content/tables/model.tex
\newcommand{\bb}[0]{\bm{b}}
\begin{table}
\centering
\small
    \caption{Different components of the runtime of a forward pass. The batch size $b$ representing the batching effect is emphasized.}
    \adjustbox{max width=\linewidth}{
    \begin{tabular}{l|c c c c c}
    \toprule
    & $T_{dm}^\linear$ & $T_{comp}^\linear$ & $T_{dm}^\attn$  & $T_{comp}^\attn$ & $T_{nw}(TP)$ \\
    \midrule
    Prefill & $\frac{2W}{\bhbm}$ & $\frac{2 \bb W s}{\flops}$ & $\frac{2\bb s(h_q + 2h_{kv}) d}{\bhbm}$ & $\frac{\bb h_q s^2 d^2}{\flops}$ & $\frac{4\bb sh_qd}{B_{ar}(TP)}$\\
    \midrule
    Decode & $\frac{2W}{\bhbm}$& $\frac{2 \bb W}{\flops}$ & $\frac{4\bb s h_{kv} d}{\bhbm}$ & $\frac{2 \bb h_q s d^2}{\flops}$ & $\frac{4\bb h_qd}{B_{ar}(TP)}$\\
    \bottomrule
    \end{tabular}
    }
    \label{tab:runtime}
\end{table}